\documentclass{article}

\usepackage[final]{neurips_data_2024}

\usepackage[utf8]{inputenc} 
\usepackage[T1]{fontenc}    
\usepackage[colorlinks=true, citecolor=blue, linkcolor=blue, urlcolor=blue]{hyperref}      
\usepackage{url}            
\usepackage{booktabs}       
\usepackage{amsfonts}       
\usepackage{nicefrac}       
\usepackage{microtype}      
\usepackage[dvipsnames]{xcolor} 
\usepackage{pdflscape}      
\usepackage{float}
\usepackage{soul}
\usepackage{pifont}     
\usepackage{lipsum}     
\usepackage{graphicx}
\usepackage{caption}
\usepackage{subcaption}
\usepackage{rotating}
\usepackage{enumitem}
\usepackage{siunitx}
\usepackage{tabularx}
\usepackage{xspace}
\usepackage{multirow}
\usepackage{changepage}
\usepackage{comment}
\usepackage{amsthm}     
\usepackage{amsmath}    
\usepackage{amssymb}    
\usepackage{mathtools}  
\usepackage[textsize=tiny]{todonotes}
\usepackage[capitalize,noabbrev]{cleveref}       

\theoremstyle{plain}

\theoremstyle{definition}

\theoremstyle{remark}

\definecolor{dandelion}{rgb}{0.94, 0.88, 0.19}

\newcommand{\abag}{Ab-Ag }

\newcommand{\VH}{\mbox{V\kern-.1667em \lower.5ex\hbox{\scriptsize H}}}
\newcommand{\VL}{\mbox{V\kern-.1667em \lower.5ex\hbox{\scriptsize L}}}

\newcommand{\CH}[1]{\mbox{C\lower.5ex\hbox{\scriptsize H}#1}}
\newcommand{\CL}{\mbox{C\kern-.0833em \lower.5ex\hbox{\scriptsize L}}}

\newcommand{\angs}[1]{\mbox{$#1$\AA}}

\newcommand{\kaggle}{\raisebox{-1.5pt}{\includegraphics[height=1.05em]{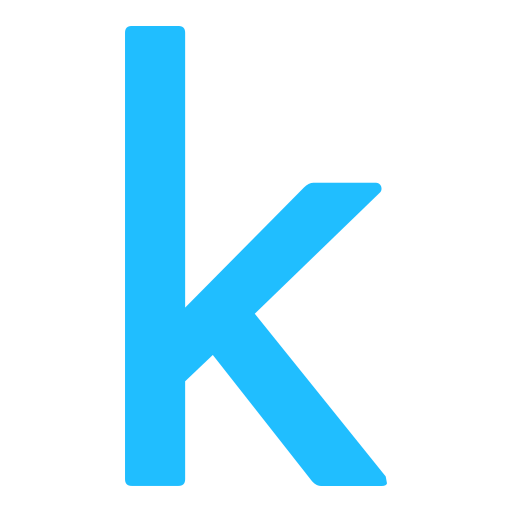}}\xspace}

\newcommand{\ccbync}{\raisebox{-1.5pt}{\includegraphics[height=1.05em]{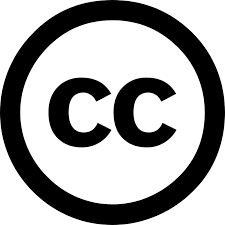}}\xspace}

\newcommand{\github}{\raisebox{-1.5pt}{\includegraphics[height=1.05em]{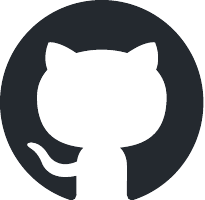}}\xspace}

\usepackage{fancyhdr}
\definecolor{ao}{rgb}{0.0, 0.5, 0.0}
\newcommand{\aurel}[1]{\textcolor{teal}{Aurelien: #1}}

\title{AbRank: A Benchmark Dataset and Metric-Learning Framework for Antibody–Antigen Affinity Ranking}

\author{%
  Chu’nan Liu$^{1,}$\thanks{A.P. and C.L. contributed equally; author order reflects relative seniority.} ,\quad
  Aurelien Pelissier$^{2,3,}$\footnotemark[1] ,\quad
  Yanjun Shao$^{3}$,\quad
  Lilian Denzler$^{1,3}$,\quad \\
  \bf 
  Andrew C.R. Martin$^{1}$,\quad
  Brooks Paige$^{4,}$\thanks{Corresponding authors: \texttt{maria.rodriguezmartinez@yale.edu} , \texttt{b.paige@ucl.ac.uk}},\quad
  Mar\'{\i}a Rodr\'{\i}guez Mart\'{\i}nez$^{3}$\footnotemark[2]\\[.1922cm]
    \small $^{1}$Structural Molecular Biology, Division of Biosciences, University College London, United Kingdom\\
    \small $^{2}$Institute of Computational Life Sciences, Zürich University of Applied Sciences (ZHAW), Switzerland\\
    \small $^{3}$Biomedical Informatics and Data Science, Yale School of Medicine, United States\\
    \small $^{4}$Centre for Artificial Intelligence, University College London, United Kingdom
}  

\begin{document}

\maketitle


\begin{abstract}
    
    Accurate prediction of antibody–antigen (Ab-Ag) binding affinity is essential for therapeutic design and vaccine development, yet the performance of current models is limited by noisy experimental labels, heterogeneous assay conditions, and poor generalization across the vast antibody and antigen sequence space. We introduce \textbf{AbRank}, a large-scale benchmark and evaluation framework that reframes affinity prediction as a pairwise ranking problem. AbRank aggregates over 380,000 binding assays from nine heterogeneous sources, spanning diverse antibodies, antigens, and experimental conditions, and introduces standardized data splits that systematically increase distribution shift, from local perturbations such as point mutations to broad generalization across novel antigens and antibodies. To ensure robust supervision, AbRank defines an \textit{m-confident ranking} framework by filtering out comparisons with marginal affinity differences, focusing training on pairs with at least an $m$-fold difference in measured binding strength.
    As a baseline for the benchmark, we introduce \textbf{WALLE-Affinity}, a graph-based approach that integrates protein language model embeddings with structural information to predict pairwise binding preferences. Our benchmarks reveal significant limitations in current methods under realistic generalization settings and demonstrate that ranking-based training improves robustness and transferability. In summary, AbRank offers a robust foundation for machine learning models to generalize across the antibody–antigen space, with direct relevance for scalable, structure-aware antibody therapeutic design.


\begin{center}
\hspace*{-2cm} 
\begin{minipage}{0.9\linewidth}
\begin{tabular}{rcl}
    \kaggle  & \textbf{\small{Data}} & \url{https://www.kaggle.com/datasets/aurlienplissier/AbRank} \\
    \github & \textbf{\small{Code}} & \url{https://github.com/biochunan/AbRank-WALLE-Affinity} \\
    \ccbync & \textbf{License\textsuperscript{\ddag}} & \url{https://creativecommons.org/licenses/by-nc/4.0/deed.en} \\
\end{tabular}
\end{minipage}
\end{center}

\renewcommand{\thefootnote}{\ddag}
\footnotetext{Details of individual licenses for each collected source are provided in Supplementary Table~\ref{tab:dataset_licenses}.}

\end{abstract}

\newpage


\section{Introduction}

Antibody (Ab) design is fundamental for therapeutic development, vaccine engineering, and diagnostics. Central to these applications, accurate antibody–antigen (Ab-Ag) affinity prediction is essential for Ab discovery, maturation, and optimization. Recent deep learning advances have improved the modeling of protein–protein interactions; however, most approaches fail to generalize across the vast and diverse Ab-Ag complex landscape, which spans thousands of Abs, highly variable Ags, and heterogeneous assay conditions. Robust, scalable affinity prediction remains a major unsolved challenge, particularly under distribution shifts involving unseen Ab and Ag scaffolds.

Most existing methods treat affinity prediction as a regression task, estimating real-valued scores from Ab and Ag sequences or structures~\citep{shan2022deep, jin2024attabseq, michalewicz2024antipasti}. While natural, this approach faces key challenges: affinity measurements are noisy, assay conditions vary widely, and values are often censored (e.g., reported only within detection limits, such as "$< 0.1$ nM"), reducing the reliability of absolute predictions. As a result, models struggle to generalize, especially to unseen Abs or Ags, or under distribution shifts. Moreover, current benchmarks often emphasize local perturbations—i.e., variants that differ by only a few mutations from the original Ab or Ag—rather than testing broader generalization across diverse Ab–Ag complexes, hence, limiting progress toward models with stronger generalization capabilities.

\begin{figure}[h!t]
\centering
    \captionsetup{width=1\linewidth}
    \includegraphics[width=0.8\linewidth]{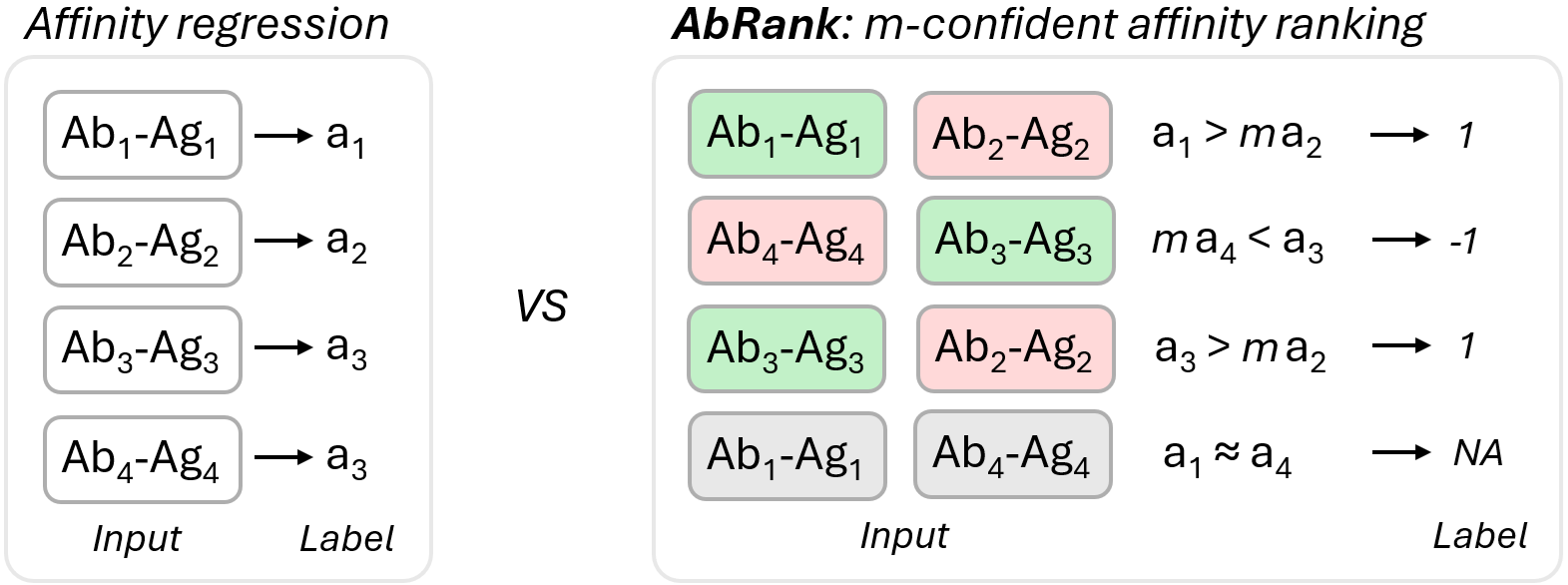}
    \caption{Regression versus pairwise ranking approaches for modeling antibody–antigen binding affinity. Pairwise ranking focuses on learning relative affinity orderings rather than predicting absolute values. AbRank only uses ranking pairs that differ by at least an $m$-fold in measured affinity, ensuring that comparisons are biologically meaningful and minimizing the impact of label noise.}
    \label{fig:Abrank}
\end{figure}

To address these challenges, we introduce \textbf{AbRank, a large-scale benchmark and evaluation framework} for Ab–Ag binding affinity prediction. AbRank reformulates affinity prediction as a pairwise ranking task, allowing models to learn relative comparisons between Ab–Ag complex pairs instead of regressing on potentially inaccurate absolute values. To reduce label noise and ensure meaningful training signals, AbRank focuses on \textit{m-confident ranking pairs}, i.e., pairs of Ab–Ag complexes with at least an $m$-fold difference in measured binding affinity. This thresholding strategy explicitly filters out ambiguous or unreliable comparisons that fall within the noise margin of experimental assays, encouraging the model to focus on clear, confident discriminations (Figure~\ref{fig:Abrank}).

We curate over 380,000 binding assays from nine public datasets, spanning diverse Abs, Ags, and experimental conditions. To assess generalization, we define standardized train/test splits that span increasing levels of sequence variation, from single amino acid changes to entirely novel Abs and Ags. We also provide predicted 3D structures for all proteins to enable structure-aware learning while maintaining scalability.

Alongside the benchmark, we introduce a metric-learning framework for affinity ranking and present \textbf{WALLE-Affinity}, a graph neural network (GNN) that integrates sequence and structure information to predict pairwise binding preferences. WALLE-Affinity serves as a strong baseline for AbRank, showing that framing affinity prediction as a ranking task improves both robustness to experimental noise and generalization across diverse Ab-Ag complexes. We benchmark WALLE-Affinity against sequence-based models, structure-aware predictors, and energy-based tools across both broad generalization and fine-grained perturbation tasks.
By establishing a large-scale, rigorously evaluated benchmark and introducing a practical alternative to regression-based modeling, AbRank provides a cornerstone for developing and fairly assessing next-generation Ab-Ag affinity predictors.

\section{Related Work}
\label{sec:related-work}

\vspace{-0.2cm}

\paragraph{Pairwise ranking over regression for robust supervision.}
Pairwise ranking losses offer a compelling alternative to regression, especially when only relative comparisons are meaningful or labels are noisy. Originally developed for information retrieval~\citep{burges2005learning, burges2006learning}, these losses model relative preferences (e.g., \( y_i < y_j \)) rather than predicting absolute scores. They have since been successfully applied  in computer vision~\citep{wang2014learning, zhang2023improving}, recommender systems~\citep{durmus2024pairwise}, and language models, particularly for fine-tuning with human feedback~\citep{qin2023large, rafailov2023direct}. In computational biology, ranking-based objectives have improved generalization to out-of-distribution data~\citep{hawkins2024likelihood} and demonstrated robustness to label imbalance~\citep{liu2024interlabelgo}.

\paragraph{Binding affinity prediction.}


Predicting binding affinity—and related tasks such as mutation impact or Ab specificity—has long been a fundamental challenge in computational biology. State-of-the-art approaches typically adopt dual-encoder architectures~\citep{shan2022deep, li2022bacpi, jin2024attabseq, bandara2024deep, zhang2025antibinder, yuan2023dg}, where separate encoders process the Ab and Ag sequences (or structures), and the resulting representations are fused via cross-attention or bi-directional attention mechanisms. These are then passed to a prediction head, often a convolutional neural network (CNN). More recently, because both Abs and Ags are proteins, graph-based representations have emerged as a natural way to capture their structural and relational properties. Such models encode residue- or atom-level interactions through message passing frameworks, with Graph Attention Networks (GATs)~\citep{velivckovic2017graph} and Graph Convolutional Networks (GCNs)~\citep{kipf2016semi} commonly used to capture both local and global structural dependencies.

Regarding input data, earlier approaches relied solely on sequence information. However, there is growing consensus that incorporating 3D structural data of the Ab–Ag complex leads to substantial performance improvements. Recent studies have demonstrated that structural and geometric features capture critical interaction details, leading to significantly higher prediction accuracy~\citep{gong2024abcan, shan2022deep, cai2024pretrainable, myung2020mcsm, zhang2020mutabind2, tubiana2022scannet}.

Models can also be categorized by their learning objectives, with most focusing on predicting the effect of mutations on binding, e.g., GearBind~\citep{cai2024pretrainable} and Bind-ddG~\citep{shan2022deep}, and only a subset directly predicting binding affinity from the full Ab–Ag complex, e.g., ANTIPASTI~\citep{michalewicz2024antipasti}. This emphasis on mutation impact is driven by two key factors: (i) the lack of a standardized framework for large-scale, general affinity prediction, and (ii) the critical importance of modeling local perturbations for applications such as affinity maturation and identifying viral escape mutations.

\paragraph{Embedding strategies for Ab and Ag sequences.}
With the rise of protein language models (PLMs), sequence-based embeddings have become foundational for encoding Abs and Ags. General-purpose models such as ESM-2~\citep{lin2022language} are widely used, while Ab-specific models such as AntiBERTy~\citep{ruffolo2021deciphering} and AbLang2~\citep{olsen2024addressing} offer domain-adapted embeddings that better capture Ab-specific features and variability. More recently, interaction-aware PLMs such as MINT~\citep{ullanat2025learning} have been introduced to model protein–protein interactions. 

These embeddings are typically used to encode nodes, e.g., residues or atoms, in a graph representation of a protein complex. Structural features are then incorporated through edge-level attributes, such as inter-residue distances, binary descriptors of spatial proximity~\citep{chunan2024asep}, and local geometric descriptors such as dihedral angles. Models such as WALLE~\citep{chunan2024asep} have shown that combining sequence-derived embeddings with rich structural priors can substantially improve downstream predictions, especially in settings where structure plays a critical role. While newer models like ESM-3~\citep{hayes2025simulating}—trained with multitask objectives incorporating structural and functional supervision—may offer more structure-aware embeddings, they remain impractical for most downstream applications requiring scalability and efficiency, due to their extremely large size (ranging from 2 to 100 billion parameters).

Modern structure prediction tools such as AlphaFold3~\citep{abramson2024accurate}, Chai-1~\citep{chai2024decoding}, and Boltz-1~\citep{wohlwend2024boltz} can generate Ab-Ag complex structures with reasonable accuracy. However, these models are highly computationally expensive, and importantly, their predicted structures often exhibit uncertainty in flexible or poorly conserved regions, areas that are critical determinants of Ab-Ag binding. This structural uncertainty can propagate into downstream models, reducing the robustness and generalization of structure-dependent methods such as GearBind~\citep{cai2024pretrainable} and ANTIPASTI~\citep{michalewicz2024antipasti}. In contrast, approaches that rely on coarse-grained graph features, e.g.,  WALLE~\citep{chunan2024asep}, are more resilient to errors in the predicted structures and can better handle structural variability, while still capturing essential spatial information.

%

\section{Problem formulation}
\label{sec:problem-formulation}

\vspace{-0.25cm}


Binding affinity prediction is typically framed as a regression task, with models trained to predict a real-valued affinity score (e.g., $K_d$, $IC_{50}$) from Ab–Ag complexes. However, this approach faces several limitations. Affinity measurements are often noisy, vary significantly across experimental protocols, and are biased toward specific Ab or Ag families, factors that jointly hinder model generalization. Moreover, many affinity values are reported only within detection limits (e.g., "$< \SI{0.1}{nM}$" or "$> \SI{100}{nM}$"), introducing censoring that violates the assumptions of standard regression losses. In some datasets, particularly those derived from deep mutational scannings, only relative binding measures are available, such as fold-changes or escape fractions (reflecting the extent to which specific mutations reduce binding), while absolute affinity values are entirely absent. These challenges reduce the reliability of absolute affinity prediction and motivate alternative objectives based on relative binding comparisons.

To address these challenges, AbRank reformulates affinity prediction as a pairwise ranking task. Rather than predicting absolute affinity scores, AbRank compares Ab–Ag complex pairs and learns which one exhibits stronger binding. This formulation aligns closely with practical applications, such as Ab design, where the goal is to prioritize candidate Abs for a given Ag target or identify escape mutations that reduce binding for a specific Ab.

\paragraph{\textit{M}-confident ranking-based formulation.}
AbRank formulates the task as an $m$-confident metric learning problem (Figure~\ref{fig:Abrank}). Each sample corresponds to an Ab–Ag complex, denoted $x_i = (g_i, b_i)$, with measured binding affinity $a_i$.
During training, the model receives pairs of complexes $(x_i, x_j)$ and a label $y_{ij}$ indicating which complex exhibits stronger binding: $y_{ij} = 1$ if $a_i > m \cdot a_j$ and $y_{ij} = -1$ if $m \cdot a_i < a_j$, where $m$ denotes the minimum required fold-change in affinity. Pairs that do not meet either condition are excluded to avoid ambiguous comparisons.
The model is then trained to minimize a margin-based ranking loss, encouraging higher predicted scores for complexes with stronger measured affinity:
%
%
\begin{align}
  \mathcal{L}(x_i, x_j, y_{ij}) = \max\left(0, -y_{ij} \cdot (f(x_i) - f(x_j)) + \text{margin}\right), 
\end{align}
where $f(x)$ is the predicted affinity score for input $x$, and $\text{margin}$ is a non-negative hyperparameter enforcing a minimum separation between pairs with known affinity orderings. 
To promote generalization, we do not restrict training to Ab–Ag complex pairs that share the same Ab or Ag. Instead, we construct ranking pairs from both intra-Ag comparisons, where we compare Ab candidates that target the same Ag, and inter-Ag comparisons, where Ab candidates target different Ags. This strategy enables the model to learn affinity orderings across a broader and more diverse contexts.

\section{Affinity ranking dataset and benchmarks}\label{sec:Ab-affinity-dataset}

\vspace{-0.3cm}

\subsection{Curating a comprehensive Ab-Ag affinity dataset}
\label{sec:curating-datasets}

\vspace{-0.1cm}


We compile Ab–Ag binding data from nine publicly available studies and datasets, covering diverse experimental conditions, target Ags, and affinity measurement metrics (Table~\ref{tab:dataset_summary}, Supplementary Figure~\ref{fig:database_count}). Some datasets focus on systematic mutational scans, e.g., AlphaSeq~\citep{engelhart2022dataset}, RBD-escape~\citep{greaney2022antibody}, while others capture diverse Ab–Ag interactions across unrelated proteins, e.g., SAbDab~\citep{dunbar2014sabdab}, SKEMPI-v2~\citep{jankauskaite2019skempi}. All retained entries include fully paired variable Ab heavy and light chains and their corresponding Ag sequences. We exclude entries lacking complete heavy–light chain pairing or missing Ag sequence data.
Furthermore, affinity is measured using heterogeneous proxies—including the dissociation constant ($K_d$), the half-maximal inhibitory concentration ($IC_{50}$), and the escape fraction ($f$)—which are not directly comparable. When multiple affinity measurements are available for a given complex, we take the median value to obtain a single representative affinity. Appendix~\ref{appendix:dataset-details} discusses transferability across different measurement types. To quantify sequence diversity, we cluster antibodies and antigens at 75\% global sequence identity. Global identity considers similarity across the entire concatenated heavy and light chain sequence, in contrast to local identity, which measures similarity over shorter matching subsequences. Identity is calculated using the Levenshtein distance~\citep{levenshtein1966binary}, normalized by the mean sequence length.

This 75\% threshold provides a balance between minimizing redundancy and preserving functional diversity, especially for highly variable Ags such as HIV Env~\citep{arrildt2012hiv, guan2013diverse}. For Abs, this threshold typically separates sequences that do not share V genes~\citep{peres2023ighv}. Clustering is also used to define data splits for downstream analyses, ensuring that similar sequences are excluded from the test set to better evaluate generalization. Full curation details, including alternative clustering thresholds, are described in Appendix~\ref{appendix:dataset-details}.


\vspace{0.2cm}
\begin{table}[h]
\vspace{-0.4cm}
    \centering
    \caption{Overview of Ab–Ag binding affinity datasets used in this study. The datasets differ in assay counts, diversity of Abs and Ags, and the types of affinity metrics reported. Affinity values include absolute measures ($K_d$, $IC_{50}$) and mutational shifts ($\Delta K_d$, escape fraction $f$). Ab75 and Ag75 indicate the number of Ab and Ag clusters at 75\% global sequence identity.}
    \vspace{0.2cm}
    \def\arraystretch{1.2}
    \resizebox{\textwidth}{!}{
    \begin{tabular}{llllr}
        \hline
        \textbf{Dataset} & \textbf{\#Assays} & \textbf{\#Ab75} & \textbf{\#Ag75} & \textbf{Affinity Data} \\
        \hline
        CATNAP~\citep{yoon2015catnap} & 74,540 & 128 & 36 & $IC_{50}$ \\
        AlphaSeq~\citep{engelhart2022dataset} & 71,834 & 3 & 1 & $K_d$ \\
        Ab-CoV~\citep{rawat2022ab} & 1,392 & 20 & 2 & $IC_{50}$, $K_d$ \\
        HER2-binders~\citep{shanehsazzadeh2023unlocking} & 762 & 1 & 1 & $K_d$ \\
        OVA-binders~\citep{goldstein2019massively} & 89 & 52 & 1 & $K_d$ \\
        SAbDab~\citep{dunbar2014sabdab} & 249 & 20 & 147 & $K_d$ \\
        AIntibody~\citep{erasmus2024aintibody} & 34 & 1 & 1 & $K_d$ \\
        SKEMPI-v2~\citep{jankauskaite2019skempi} & 935 & 11 & 21 & $K_d$, $\Delta\log(K_d)$ \\
        RBD-escape~\citep{greaney2022antibody} & 192,559 & 19 & 1 & $f \sim \Delta\log(K_d)$ \\
        \hline
        \textbf{Our curated dataset} & \textbf{381,713} & \textbf{237} & \textbf{190} & -- \\
        \hline
    \end{tabular}
    }
    \label{tab:dataset_summary}
\end{table}
\def\arraystretch{1.0}

\vspace{-0.2cm}

\subsection{Generating Ag and Ab structures}

\vspace{-0.2cm}

We generate 3D structures for all Abs and Ags in our dataset. Ab structures are generated using IgFold~\citep{ruffolo2023fast}, a model optimized explicitly for Ab modeling. For Ags, we used \mbox{Boltz-1}~\citep{wohlwend2024boltz} to directly predict full-chain structures. To enable large-scale processing, we prioritize structure prediction methods that support fast, single-sequence inference, favoring computationally efficient models over more resource-intensive alternatives such as \mbox{AlphaFold-3} (AF3)~\citep{abramson2024accurate}. Although these choices may introduce greater structural noise, they offer a favorable trade-off between speed and accuracy, making them well-suited for high-throughput applications.
%


\subsection{Pairwise dataset construction}
\label{sec:database-split}

\vspace{-0.2cm}


We construct a \textit{paired dataset} by sampling affinity measurements with a known ordering, defined by a minimum fold-change $m$ in affinity (i.e., $\Delta \log_{10}(K_d) > m$). Unless otherwise noted, we use $m = 1$, corresponding to a 10-fold difference in binding affinity. For illustration, a single amino acid replacement—particularly at the binding interface—can shift the dissociation constant ($K_d$) by factors ranging from 10-fold to over 10,000-fold~\citep{foote1995kinetic, boder2000directed}. In our dataset, observed $K_d$ values span from $10^{-4} , \text{nM}$ to $10^{4} , \text{nM}$, covering a dynamic range of up to $10^8$-fold in binding affinity (Supplementary Figure~\ref{fig:database_affinity}).
%
This threshold ensures that pairs reflect confidently distinguishable differences—an important consideration given the inherent noise and variability in experimental affinity measurements. Within the studied range, $K_d$ values obtained through surface plasmon resonance (SPR)—a widely used, label-free technique for quantifying real-time biomolecular interactions—are generally robust. However, our dataset also includes measurements from biolayer interferometry (BLI) and enzyme-linked immunosorbent assays (ELISA), techniques which are generally noisier and more sensitive to variations in experimental conditions across platforms, potentially reducing measurement precision~\citep{thinn2025competitive}. By focusing on larger fold-changes, we mitigate the effect of noise and encourage the model to learn from functionally meaningful distinctions in binding strength. In contrast, smaller fold changes, e.g., 1.5- to 2-fold, are more susceptible to batch effects and assay-specific variability.
%
While \( K_d \) and \( IC_{50} \) have different biophysical interpretations, we treat them as interchangeable for ranking purposes. This is based on the assumption that sufficiently large differences (e.g., \( \geq 10\times \)) in either metric generally reflect substantial changes in binding affinity~\citep{cer2009ic, swinney2011molecular}. 


Rather than exhaustively using all available data, we focus on constructing a dataset that emphasizes balance and diversity. To reduce sampling bias and improve coverage, we sample affinity pairs as evenly as possible across Ab and Ag clusters (Ab75 and Ag75), i.e., defined using 75\% global sequence identity. We include both \emph{intra-Ab-Ag} pairs (within the same Ab–Ag group) and \emph{inter-Ab-Ag} pairs (across distinct groups), allocating 25\% and 75\% of the dataset to each, respectively. This split promotes generalization across molecular contexts while retaining sensitivity to local sequence variations. It also increases the diversity and variance of ranking comparisons, reducing the risk of overfitting to oversampled Ab–Ag complexes.

\begin{figure}[h!t]
    \centering
    \begin{minipage}[t]{0.495\textwidth}

    After clustering and filtering, we obtain affinity data for approximately \emph{500 unique Ab–Ag complex groups}, i.e., unique Ab75–Ag75 cluster pairs. The distribution of binding assays count per complex group is highly imbalanced: 139 complexes have more than 10 measurements, 64 exceed 100, and 10 have over 1,000. To mitigate this imbalance in the intra-group setting, we subsample \emph{at most 1,000 pairwise comparisons per complex group}, resulting in approximately 50k intra-group comparisons with >10-fold affinity differences. For inter-group comparisons, we sample with a weight proportional to the square root of the size of the clusters, hence reducing the sampling of groups that have only a few binding assays. In this setting, we first select two distinct Ab–Ag clusters, then draw one complex at random from each.
    \end{minipage}%
    \hfill
    \begin{minipage}[t]{0.477\textwidth}
    \vspace{-0.42cm}
    \centering
    \includegraphics[width=\linewidth]{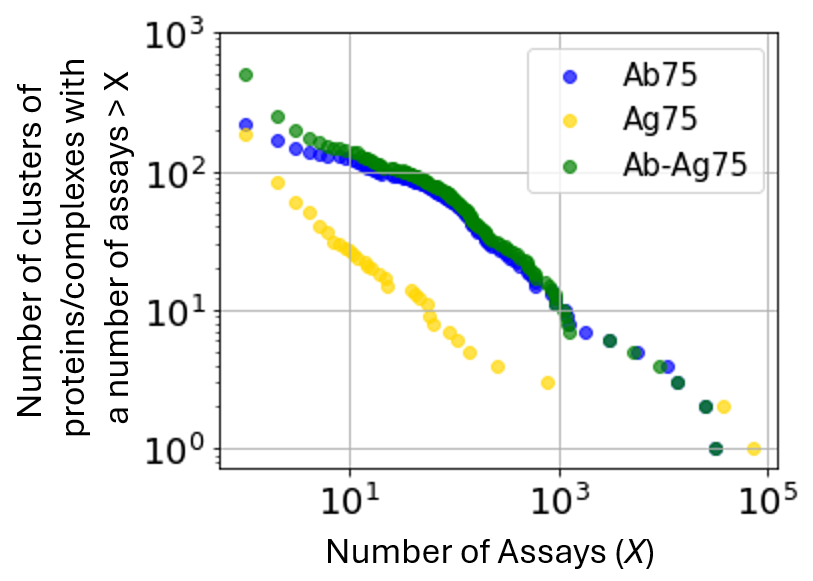}
    \captionof{figure}{ \small Cumulative number of Abs, Ags, and Ab–Ag complexes with at least a given number of binding assays on our curated dataset. Clusters are defined using a 75\% global sequence identity threshold.}
    \label{fig:Ab-Agcount}
    \end{minipage}
    \vspace{-0.3cm}
\end{figure}

\subsection{Test set and benchmarking}

\vspace{-0.2cm}

To prevent data leakage into baseline models trained on known complex structures, such as GearBind~\citep{cai2024pretrainable} and ANTIPASTI~\citep{michalewicz2024antipasti}, we select Ab–Ag complexes whose antibody heavy and light chains each share no more than 75\% sequence identity with any antibody in SAbDab~\citep{dunbar2014sabdab} (accessed May 1, 2025). As a result, all complexes involving antibodies from SKEMPI v2 and SAbDab are excluded, avoiding overlap with training data used by many existing methods and eliminating dependence on known crystallized structures from the Protein Data Bank (PDB), which are disproportionately represented in these datasets.
From the filtered set, we sample 362 Ab–Ag complexes spanning 45 distinct Ab75–Ag75 groups, ensuring balanced representation across Ags. The selected complexes include binders targeting SARS-CoV-2, HIV, and ovalbumin (OVA), reflecting realistic use cases where the true Ab–Ag complex structure is unavailable.

\paragraph{Two-tier benchmarking by Ab-Ag complex pair similarity.}

We structure our evaluation around two distinct settings. The first, referred to as the \textbf{Unrelated Complex Benchmark}, evaluates the model's ability to generalize across highly diverse and unrelated Ab–Ag complexes by ranking affinities for arbitrary pairs with no constraints on shared sequence or structural similarity. This setting does not aim to capture affinity maturation but instead tests out-of-distribution generalization across the broader Ab–Ag space. The second, referred to as the \textbf{Local Perturbation Benchmark}, targets fine-grained ranking within Ab–Ag variants that belong to the same group—defined as having over 75\% sequence similarity for both Abs and Ags. This second benchmark reflects a therapeutically relevant scenario, supporting the evaluation of antibody optimization strategies (e.g., affinity maturation) as well as the assessment of viral escape mutations.

\paragraph{Three-tier training split by Ab–Ag generalization.}
Affinity datasets are highly imbalanced, not only in measurement counts but also in their Ab and Ag distributions. Some Ags, such as SARS-CoV-2 spike variants, are heavily overrepresented due to mutational scanning studies, while others appear sparsely or in only a handful of complexes. To evaluate generalization under varying levels of sequence distribution shift, we define three controlled training splits based on Ab–Ag composition. All splits adhere to a key constraint: during evaluation, the model must rank pairs of Ab–Ag complexes not seen during training. This ensures that each comparison reflects some form of generalization, while allowing precise control over what is considered "unseen"—whether at the level of the specific complex, the Ab sequence, the Ag sequence, or both.

\begin{itemize} 
    \item \textbf{Balanced split}: A baseline setting where each test comparison involves two Ab–Ag complexes not seen during training. However, unseen complexes may be similar,  e.g., differing by only a few amino acids, to those in the training set. This split primarily tests generalization to small sequence variations.
    \item \textbf{Hard Ab split}: Each test comparison involves two Abs with less than 75\% global sequence identity to any Ab seen during training. Ags may still overlap across training and test sets. This split tests generalization across the Ab sequence space. 
    \item \textbf{Hard Ag split}: Each test comparison involves two Ags with less than 75\% sequence identity to any training Ag. Abs may still overlap across training and test sets. However, because it is very uncommon for a given Ab to bind multiple highly dissimilar Ags~\citep{corti2013broadly}, this split typically also involves unseen Abs. This setting tests generalization to new antigenic targets and, indirectly, to novel Ab–Ag combinations.
    %
\end{itemize}

These stratified splits allow us to dissect the contribution of Ab and Ag variation to generalization performance and to evaluate the model’s robustness under increasing levels of distributional shift.

\section{WALLE-Affinity}\label{sec:WALLE_affinity_ranking}

\vspace{-0.2cm}

\textbf{Model architecture.} WALLE~\citep{chunan2024asep} is a graph-based method for Ab-specific epitope prediction, trained on a curated dataset of 1,723 Ab–Ag complexes. It predicts an Ab’s binding site on an Ag surface and serves as the foundation for our model. We adapt the pretrained WALLE model to our task, which requires predicting the binding affinity of Ab–Ag pairs. WALLE consists of a graph encoder and a decoder. In our adaptation, we retain the graph encoder and replace the decoder with a regressor head that outputs an affinity score. For graph construction, we follow \citep{chunan2024asep} and use the best-performing combination of protein embeddings: AntiBERTy~\citep{ruffolo2021deciphering} for Abs and ESM2-650M~\citep{lin2022language} for Ags.
To leverage the pretrained WALLE model, we initialize its encoder blocks using the original WALLE checkpoint and replace the decoder with a regression module. This module applies global mean pooling to the Ab and Ag graph embeddings, concatenates the resulting vectors, and maps them to a scalar affinity score via a linear layer.


\textbf{Training details.}  
We implement a dataloader that samples batches of Ab–Ag complex pairs. Each pair consists of two \abag graphs. We followed the same procedure to construct graphs from individual Ab or Ag structures as reported in AsEP \citep{chunan2024asep}, except that edges are defined using a distance threshold of \angs{10} between $C_\alpha$ atoms. A pair is labeled as positive if the first complex exhibits at least a 10-fold higher binding affinity than the other, and negative if it exhibits at least a 10-fold lower affinity (i.e., the inverse). Pairs with less than a 10-fold difference in affinity (i.e., within the margin) are excluded from training. Each batch is shuffled such that half the pairs are positive and half are negative. The model is trained with the marginal pairwise ranking loss, using early stopping with a patience of 10 steps and a margin parameter of 0.1. We use a learning rate of \(1 \times 10^{-5}\) with a \texttt{linear\_warmup\_and\_cosine\_annealing} scheduler.

\section{Experiments and Discussion}\label{sec:Discussion}

\vspace{-0.25cm}

We evaluate our model, WALLE-Affinity, on both an unrelated complex benchmark and a local perturbation benchmark. Results are reported across the three training splits introduced in Section~\ref{sec:database-split}, which progressively increase generalization difficulty by removing from the training shared Abs and/or Ags between the training and test sets (Table~\ref{tab:main-results}). We compare versions of WALLE-Affinity trained using both regression and pairwise ranking objectives. In the regression approach, the model is trained to predict continuous affinity values for individual Ab–Ag complexes; test-time rankings are then derived by comparing predicted scores across candidate pairs.
We benchmark our model against a broad range of baselines spanning three categories: (i) sequence-only approaches such as ESM-2~\citep{lin2022language} and Mint~\citep{ullanat2025learning}, a fine-tuned ESM-2 model for protein–protein interactions; (ii) state-of-the-art affinity prediction methods such as ANTIPASTI~\citep{michalewicz2024antipasti} and GearBind~\citep{cai2024pretrainable}; and (iii) energy-based computational tools like PBEE~\citep{chaves2025estimating} and FoldX~\citep{delgado2019foldx}. For methods that require an Ab–Ag complex structure, we generate the Ab-Ag complexes using Boltz1~\citep{wohlwend2024boltz} and AlphaFold3 (AF3)~\citep{abramson2024accurate}.  The implementation details of each baselines can be found in Appendix~\ref{SI:baseline}.
Performance for each method is reported as the area under the ROC curve (AUC), summarized in Table~\ref{tab:main-results}.

\def\arraystretch{1.2}
\begin{table}[t]
    \vspace{-0.6cm}
    \centering
    \captionsetup{width=1.17\textwidth}
    \caption{Area under the ROC curve (AUC) for each method across the three benchmark splits. The first three columns present results on the \textbf{Unrelated Complex Benchmark}, which assesses ranking performance across diverse, unrelated Ab–Ag complexes. The last three columns correspond to the \textbf{Local Perturbation Benchmark}, evaluating fine-grained affinity ranking within Ab–Ag complexes belonging to the same 75\% similarity group—that is, differing by only a few mutations. This setting reflects scenarios such as affinity maturation and viral escape. WALLE-Affinity is compared to sequence-based, structure-aware, and energy-based baselines. Methods prefixed with \textbf{Boltz1+} and \textbf{AF3+} operate on Boltz1 and AF3-predicted Ab–Ag complex structures, modeling a realistic scenario where the true complex structure is unknown and must be inferred computationally. Methods including ANTIPASTI, GearBind, PBEE, and FoldX were originally trained or parametrized on datasets such as SAbDab, SKEMPI, or other crystallographically resolved protein–protein interaction affinity data; in our benchmark, this corresponds most closely to the Hard Ab setting. Additionally, GearBind is specifically designed to compare affinity between closely related mutants and is therefore not applicable to the Unrelated Complex Benchmark, denoted by N/A in the table. The best AUC scores within each benchmark split are highlighted in bold separately for embedding-based and structure-based categories.}
    \vspace{0.15cm}
    \label{tab:main-results}
    \makebox[\textwidth][c]{%
    \resizebox{1.17\textwidth}{!}{%
        \begin{tabular}{lcccccc}
            \toprule
            \multirow{2}{*}{\textbf{Model}} & \multicolumn{3}{c}{\textbf{Unrelated Complex Benchmark}} & \multicolumn{3}{c}{\textbf{Local Perturbation Benchmark}} \\
            \cmidrule(lr){2-4} \cmidrule(lr){5-7}
            & \textbf{Balanced} & \textbf{Hard Ab} & \textbf{Hard Ag} & \textbf{Balanced} & \textbf{Hard Ab} & \textbf{Hard Ag} \\
            \midrule
            \midrule
            \textbf{WALLE-Affinity (ranking)} & \textbf{0.866} & \textbf{0.763} & 0.746 & 0.671 & \textbf{0.581} & \textbf{0.637} \\
            WALLE-Affinity (regression) & 0.833 & 0.753 & 0.612 & \textbf{0.685} & 0.552 & 0.610 \\
            \midrule
            ESM-2 + AntiBERTy (ranking)~\citep{ruffolo2021deciphering} & 0.761 & 0.719 & 0.758 & 0.595 & 0.516 & 0.532 \\
            ESM-2 + AntiBERTy (regression)~\citep{ruffolo2021deciphering} & 0.836 & 0.740 & \textbf{0.814} & 0.572 & 0.467 & 0.478 \\
            \midrule
            Mint (ranking)~\citep{ullanat2025learning} & 0.775 & 0.741 & 0.688 & 0.625 & 0.538 & 0.497 \\
            Mint (regression)~\citep{ullanat2025learning} & 0.456 & 0.429 & 0.428 & 0.446 & 0.398 & 0.488 \\
            \midrule
            \midrule
            Boltz1+ANTIPASTI~\citep{michalewicz2024antipasti} & \multicolumn{3}{c}{0.400} & \multicolumn{3}{c}{0.524} \\
            AF3+ANTIPASTI~\citep{michalewicz2024antipasti} & \multicolumn{3}{c}{0.459} & \multicolumn{3}{c}{\textbf{0.597}} \\
            Boltz1+GearBind~\citep{cai2024pretrainable} & \multicolumn{3}{c}{N/A} & \multicolumn{3}{c}{0.531} \\
            AF3+GearBind~\citep{cai2024pretrainable} & \multicolumn{3}{c}{N/A} & \multicolumn{3}{c}{0.573} \\
            \midrule
            Boltz1+PBEE~\citep{chaves2025estimating} & \multicolumn{3}{c}{0.364} & \multicolumn{3}{c}{0.469} \\
            AF3+PBEE~\citep{chaves2025estimating} & \multicolumn{3}{c}{0.362} & \multicolumn{3}{c}{0.480} \\
            Boltz1+FoldX~\citep{delgado2019foldx} & \multicolumn{3}{c}{0.504} & \multicolumn{3}{c}{0.517} \\
            AF3+FoldX~\citep{delgado2019foldx} & \multicolumn{3}{c}{\textbf{0.689}} & \multicolumn{3}{c}{0.585} \\
            \bottomrule
            \end{tabular}
        }%
    }
\vspace{-0.5cm}

\end{table}


\def\arraystretch{1.0}

Overall, WALLE-Affinity achieves state-of-the-art performance across most benchmark settings, establishing itself as a strong candidate for Ab–Ag affinity prediction. Notably, it strikes a compelling balance between predictive accuracy and inference speed (Table~\ref{tab:speed-results}). Despite relying solely on the individual structures of antibodies and antigens—without requiring full Ab–Ag complexes—WALLE-Affinity consistently outperforms or matches the accuracy of substantially slower methods like ANTIPASTI, GearBind, PBEE, and FoldX. These baselines were originally trained and evaluated on crystallographic Ab–Ag complexes and are highly sensitive to small structural variations introduced during complex structure prediction, which likely contributes to their degraded performance when applied to predicted inputs.

Another key insight is the clear advantage of the ranking-based objective over direct regression, observed across both WALLE-Affinity and Mint. This is especially pronounced in the unrelated complex benchmark, where the ranking variant of WALLE-Affinity achieves an AUC improvement of over 12 points compared to its regression counterpart under the Hard Ag setting. These results suggest that learning from pairwise comparisons more effectively captures relative affinity relationships, particularly when generalizing to previously unseen antigens or antibodies.

As expected, performance declines as benchmark difficulty increases, with the Hard Ag setting proving most challenging in the unrelated benchmark. One exception is the ESM-2 + AntiBERTy regression model, which performs best in this setting. Interestingly, this may be partially explained by differences in training set size (Supplementary Table~\ref{tab:dataset-count}): the Hard Ag split contains significantly fewer training samples than the Hard Ab split. Models like ESM-2 and Mint, which use fixed-size max-pooled embeddings, may be better suited to small training sets due to their reduced input dimensionality. In contrast, WALLE-Affinity encodes amino acid–level embeddings, providing richer representations at the cost of higher sample complexity.
Paradoxically, performance on the Hard Ab split in the local perturbation benchmark is often lower than on Hard Ag, despite having access to a larger training set. This may be due to overfitting on highly specific local perturbations, which do not generalize well to unseen antibodies. In such cases, more training data—if not sufficiently diverse—can actually hurt generalization.

Finally, we observe that both WALLE-Affinity and embedding-based models perform worse on the local perturbation benchmark than in the unrelated complex setting. This benchmark requires distinguishing subtle differences between closely related Ab–Ag complexes that vary by only a few mutations. These small changes can lead to large, non-linear shifts in binding affinity, making accurate prediction particularly challenging. Errors in structural prediction or representation are amplified, and affinity changes are often context-dependent and poorly transferable across examples. For instance, Mint occasionally exhibits inverse correlations between predicted and actual affinity changes under the Hard Ab split. These findings underscore the difficulty of local perturbation ranking and the need for models that can capture fine-grained, context-specific biophysical effects.


\def\arraystretch{1.1}
\begin{table}[t]
    \vspace{-0.6cm}
    \centering
    \captionsetup{width=1.02\textwidth}
    \caption{Inference time per Ab–Ag complex for each method, measured using a standard single GPU (e.g., NVIDIA RTX 3060). Mint operates solely on sequence and does not require structural input. WALLE-Affinity requires structures of both the Ab (typically generated with ImmuneFold) and the Ag (from crystal structures or predicted with Boltz1). All other methods require full complex structure prediction, which can be inferred using models such as Boltz1 or AF3. Note that AF3 is not feasible on an RTX 3060 due to its high memory and computational requirements, and typically requires a high-end GPU such as an NVIDIA A100 or RTX 4090. *The reported AF3 runtime is based on inference from the AlphaFold Server web interface.}

    \label{tab:speed-results}
    \makebox[\textwidth][c]{%
    \resizebox{1.02\textwidth}{!}{%
        \begin{tabular}{llr}
            \toprule
            \textbf{Model} & \textbf{Inference Speed} & \textbf{Main processing steps} \\
            \midrule
            IgFold~\citep{ruffolo2023fast} & $\sim$20sec & Ab Structure generation\\
            Boltz-1~\citep{wohlwend2024boltz} & $\sim$1min & MSA + Structure generation\\
            AF3~\citep{abramson2024accurate} & $\sim$30min* & Structure generation\\
            \midrule
            ESM-2~\citep{ruffolo2021deciphering} & $\sim$1sec & Embeddings\\
            Mint~\citep{ullanat2025learning} &$\sim$1sec & Embeddings\\
            \textbf{WALLE-Affinity} & $\sim$10sec & Graph construction + Embeddings\\ 
            FoldX~\citep{delgado2019foldx} & $\sim$5min & Structure refinement \\
            GearBind~\citep{cai2024pretrainable} & $\sim$5min & Structure refinement + Graph construction\\
            ANTIPASTI~\citep{michalewicz2024antipasti} & $\sim$15min & Structure refinement + Normal mode\\
            PBEE~\citep{chaves2025estimating} & $\sim$20min & Structure refinement (Rosetta)\\
            \bottomrule
            
        \end{tabular}
    }%
    }
\vspace{-0.3cm}
\end{table}
\def\arraystretch{1.0}

\section{Conclusion}

\vspace{-0.3cm}
In this work, we introduce a unified benchmark dataset for antibody-antigen (Ab-Ag) affinity prediction, designed to standardize evaluation as a ranking problem across two important scenarios: generalization to unseen antibodies and/or antigens and sensitivity to local mutations. We demonstrate that formulating the affinity prediction as a pairwise ranking task mitigates challenges arising from noisy and heterogeneous experimental labels. Our benchmark dataset also provides well-defined training and test splits, enabling consistent comparisons between different computational approaches. 
As a reference baseline, we propose WALLE-Affinity, which leverages structural graph representations and pretrained protein language models. WALLE  achieves strong performance across diverse evaluation conditions, outperforming several structure-based and energy-based baselines, while remaining efficient and scalable in real-world inference settings.
Our findings highlight the effectiveness of combining structural graph representations with pretrained protein language models and underscore the advantages of ranking-based supervision for learning meaningful affinity patterns. While promising, several challenges remain. First, variability across affinity measurement types (e.g., $K_d$ vs. $IC{50}$) introduces uncertainty in the learning process. Second, by transforming computationally inferred atomistic representations of antibodies and antigens into graphs, WALLE-Affinity may help absorb structural inaccuracies, particularly common in antibodies and, more broadly, in proteins with flexible domains. However, the extent to which this abstraction mitigates structural inaccuracies remains an open question that warrants further investigation. 
Future work will focus on expanding structural annotations, refining pairwise sampling strategies, and improving interpretability to support downstream applications such as  therapeutic antibody design and immune escape modeling.


\newpage
\bibliography{ref}
\bibliographystyle{icml2022}


\appendix
\renewcommand{\thefigure}{S\arabic{figure}}   
\renewcommand{\theHfigure}{S\arabic{figure}}  
\setcounter{figure}{0}  
\renewcommand{\thetable}{S\arabic{table}}   
\renewcommand{\theHtable}{S\arabic{table}}  
\setcounter{table}{0}  


\section{Dataset Curation Details}
\label{appendix:dataset-details}

\subsection{Datasets}

\begin{figure}[h!t]
\centering
    \captionsetup{width=1\linewidth}
    \includegraphics[width=0.8\linewidth]{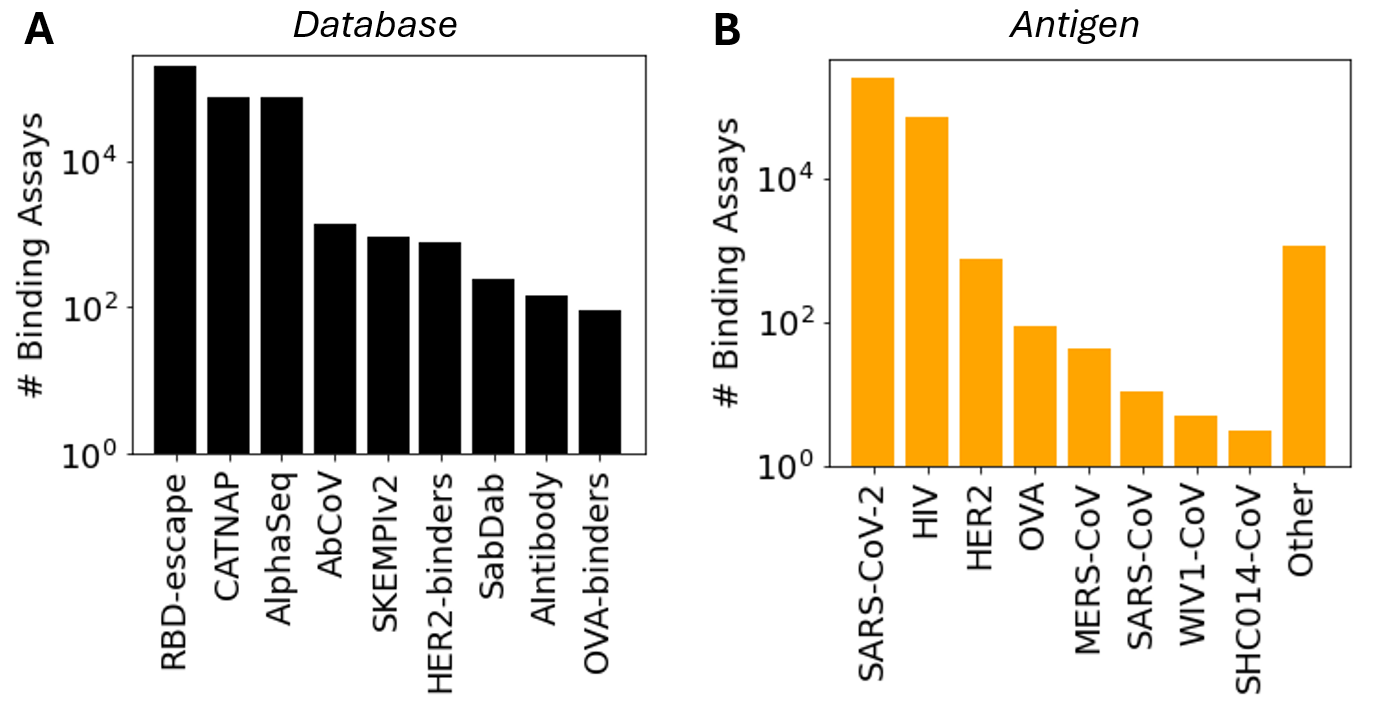}
    \caption{Distribution of the number of binding affinity assays in the AbRank dataset. (A) Number of assays per source database. (B) Number of assays per target antigen.}
    \label{fig:database_count}
\end{figure}

Our combined dataset is not only large but also diverse, spanning thousands of distinct antibodies and antigens. Despite its scale, it exhibits substantial class imbalance, both across and within datasets (see Figure~\ref{fig:database_count}). Below, we briefly describe the data curation steps for each source.

\textbf{CATNAP}~\citep{yoon2015catnap}: A dataset of HIV-neutralizing antibodies reporting $IC_{50}$ values against a wide panel of viral antigens, including a large diversity of HIV variants. We downloaded the dataset directly from the official CATNAP website (accessed May 1, 2025).

\textbf{AlphaSeq}~\citep{engelhart2022dataset}: A systematic mutagenesis dataset measuring $K_d$ values for antibody variants binding to a single linear SARS-CoV-2 peptide located outside the receptor-binding domain (RBD). We downloaded the dataset from \url{https://zenodo.org/records/5095284}, removed control binders, and deduplicated repeated measurements by taking the median affinity.

\textbf{Ab-CoV}~\citep{rawat2022ab}: A dataset aggregating antibody binding measurements against a range of coronaviruses, including SARS-CoV, SARS-CoV-2, MERS-CoV, and related bat coronaviruses. We included all entries reporting either $IC_{50}$ or $K_d$ values and involving single-chain protein antigens. The dataset was downloaded from the official project website.

\textbf{HER2-binders}~\citep{shanehsazzadeh2023unlocking}: Reports $K_d$ measurements for a large panel of antibody variants generated via directed mutagenesis, all targeting the HER2 antigen. We filtered for single-chain antigen targets and retained only entries with valid affinity values. Data was obtained from \url{https://github.com/AbSciBio/unlocking-de-novo-antibody-design/tree/main}.

\textbf{SAbDab}~\citep{dunbar2014sabdab}: A structural database of antibody–antigen complexes. We accessed the database on May 1, 2025, and downloaded all available entries, then filtered for those with associated $K_d$ values. To reduce redundancy, we removed duplicate Ab–Ag complexes (i.e., entries with identical sequences) and, for each unique pair, retained the median $K_d$ value and selected one representative structure at random from the available PDBs.

\textbf{AIntibody Challenge}~\citep{erasmus2024aintibody}: A dataset of antibodies binding to the SARS-CoV-2 RBD, released as part of the AIntibody Challenge. The dataset was downloaded directly from the official challenge website.

\textbf{OVA-binders}~\citep{goldstein2019massively}: A dataset of antibody sequences specific to the model antigen ovalbumin (OVA), generated through high-throughput single-cell B-cell receptor sequencing. Data was obtained from the supplementary materials of the publication, and we filtered for paired sequences with measured affinity values.

\textbf{SKEMPI 2.0}~\citep{jankauskaite2019skempi}: Contains experimental binding affinity measurements for mutant and wild-type protein–protein complexes, including a subset of antibody–antigen interactions. We collected entries with available $K_d$ values and retained antibody–antigen pairs for which structural data was available. Due to missing residues in some crystal structures, we aligned the reported mutation positions to the full antibody/antigen sequences in order to correctly map them onto the graph representations used by WALLE-Affinity. The dataset was obtained from the official SKEMPI website.

\textbf{RBD-escape}~\citep{greaney2022antibody}: Collecting measurement from multiple studies of Deep mutational scanning (DMS). It provides the escape fraction ($f$) of SARS-CoV-2 RBD mutants against a large panel of antibodies. Despite involving a single antigen, the dataset offers high antibody diversity. We used the processed dataset available at \url{https://github.com/ericzwang/RBD_AB/tree/main}. We convert the escape fraction $f$ to $\Delta \log(K_d)$ as described in Supplementary Section~\ref{SI:DGvsKD}.

\subsection{Grouping Antibodies and Antigens by sequence similarity}
We cluster all unique antibody and antigen sequences from our curated dataset. For antibodies, we define similarity by concatenating the heavy and light variable regions. We computed pairwise distances using the normalized Levenshtein distance (edit distance divided by the average sequence length of the pair). We apply agglomerative hierarchical clustering with \texttt{single} linkage, which ensures that the minimum distance between any two sequences in different clusters exceeds the specified threshold. We perform clustering at 90\%, 75\%, and 50\% similarity thresholds and report the resulting statistics in Table~\ref{tab:data_cluster}.

We decide to cluster antibody and antigen sequences at 75\% global identity, a threshold motivated by the extreme sequence diversity of HIV and other viral antigens. In particular, the HIV Env glycoprotein exhibits pairwise identities as low as 60–70\% across known strains. Clustering at 90\% (as in UniRef90) leads to excessive fragmentation, especially in pan-viral analyses. The 75\% threshold reflects a practical compromise—capturing major functional variation while reducing redundancy. Similar identity ranges (75–80\%) have been used in viral genomics to delineate strain-level diversity~\citep{hassan2017defining}.

Interrestingly, our data is skewed by extreme sequence diversity is the HIV Env protein. Although Env variants share a common function—mediating host cell entry via CD4 receptor binding—they exhibit remarkable sequence divergence. Across all known HIV-1 variants, less than 60\% of Env amino acid positions are conserved, with pairwise sequence identities frequently dropping below 70\%. This diversity is driven by relentless immune pressure, particularly from neutralizing antibodies, which fuels rapid escape mutations. Despite this variability, the virus maintains functional envelope proteins, highlighting its capacity to tolerate substantial structural and sequence alterations while preserving infectivity.

\vspace{0.2cm}
\def\arraystretch{1.2}
\begin{table}[h]
    \centering
    \captionsetup{width=1.15\textwidth}
    \caption{Overview of antibody–antigen binding affinity datasets used in this study. For each dataset, we report the number of assays and the cluster diversity of antibodies and antigens at different global sequence identity thresholds (90\%, 75\%, and 50\%), computed following the UniRef protocol~\citep{suzek2015uniref}. This reflects the diversity of unique antibody and antigen sequences captured at increasingly stringent similarity cutoffs.}
    \label{tab:data_cluster}
    \vspace{0.15cm}
    \makebox[\textwidth][c]{%
    \resizebox{1.15\textwidth}{!}{%
    \begin{tabular}{|l|c||c|c||c|c||c|c|}
        \hline
        \textbf{Dataset} & \textbf{\#Assays} & \textbf{\#Ab90} & \textbf{\#Ag90} & \textbf{\#Ab75} & \textbf{\#Ag75} & \textbf{\#Ab50} & \textbf{\#Ag50} \\
        \hline
        CATNAP~\citep{yoon2015catnap} & 74,540 & 381 & 1195 & 128 & 36 & 2 & 3 \\
        AlphaSeq~\citep{engelhart2022dataset} & 71,834 & 3 & 1 & 3 & 1 & 3 & 1 \\
        Ab-CoV~\citep{rawat2022ab} & 1,392 & 708 & 3 & 20 & 2 & 1 & 2 \\
        HER2-binders~\citep{shanehsazzadeh2023unlocking} & 758 & 1 & 1 & 1 & 1 & 1 & 1 \\
        OVA-binders~\citep{goldstein2019massively} & 89 & 88 & 1 & 52 & 1 & 1 & 1 \\
        SAbDab~\citep{dunbar2014sabdab} & 249 & 177 & 167 & 20 & 147 & 1 & 137 \\
        AIntibody~\citep{erasmus2024aintibody} & 142 & 14 & 1 & 1 & 1 & 1 & 1 \\
        SKEMPI-v2~\citep{jankauskaite2019skempi} & 935 & 33 & 23 & 11 & 21 & 3 & 20 \\
        RBD-escape~\citep{greaney2022antibody} & 192,559 & 1049 & 1 & 19 & 1 & 1 & 1 \\
        \hline
        \textbf{Our curated dataset} & \textbf{342,498} & \textbf{2324} & \textbf{1374} & \textbf{237} & \textbf{188} & \textbf{7} & \textbf{144} \\
        \hline
    \end{tabular}
    }%
    }
\end{table}
\def\arraystretch{1.0}

\subsection{Relationship Between $K_d$, $\Delta\Delta G$ and Escape Fraction}

\label{SI:DGvsKD}

To integrate heterogeneous binding data sources, we examine the relationships between thermodynamic free energy changes ($\Delta\Delta G$), equilibrium binding constants ($K_d$), and functional escape fractions from deep mutational scanning (DMS). Our goal is to place these quantities on a common conceptual axis: the degree to which a mutation perturbs antibody–antigen binding.

\paragraph{From $K_d$ to $\Delta G$.}
The free energy of binding is related to the dissociation constant $K_d$ through the thermodynamic identity:
\[
\Delta G = RT \log K_d,
\]
where $R$ is the gas constant (1.987 cal/mol·K) and $T$ is the temperature in Kelvin (typically 298 K). This equation allows for direct conversion between affinity values and energy terms.

\paragraph{From $\Delta G$ to $\Delta\Delta G$.}
Mutational effects on binding are commonly expressed as the change in free energy between the mutant and the reference (wild-type) complex:
\[
\Delta\Delta G = \Delta G_{\text{mut}} - \Delta G_{\text{ref}} = RT \log \left( \frac{K_d^{\text{mut}}}{K_d^{\text{ref}}} \right).
\]
Here, positive values of $\Delta\Delta G$ indicate reduced binding affinity (destabilization), while negative values indicate improved binding. The SKEMPI dataset provides curated measurements of $\Delta\Delta G$ for a wide range of point mutations in protein–protein interfaces, including antibody–antigen complexes.

\paragraph{Escape Fraction as a Functional Proxy.}
In DMS datasets, the \textit{escape fraction} $f$ measures how strongly a mutation reduces recognition by a specific antibody, with values ranging from 0 (no escape) to 1 (complete escape). Under the assumption that escape reflects loss of binding due to an increased $K_d$, we can relate $f$ to $\Delta\Delta G$ using a simplified thermodynamic model. Assuming 1:1 Langmuir binding~\citep{mollerup2008review} and high-affinity wild-type antibodies, we derive:

\[
\Delta\Delta G \approx RT \log \left( \frac{1}{1 - f} \right)
\]

where $R$ is the gas constant and $T$ the temperature. This expression shows that as escape fraction increases, the estimated destabilization ($\Delta\Delta G$) grows non-linearly. Notably, because $f$ only captures reductions in binding (not improvements), it saturates near zero for affinity-enhancing mutations. However, this limitation is acceptable in practice, since wild-type antibodies in DMS assays typically exhibit very high affinity, leaving little room for further improvement.

\subsection{Relationship Between $K_d$ and $IC_{50}$}

\label{SI:IC50vsKD}

While both $K_d$ and $IC_{50}$ aim to quantify molecular binding affinity, they arise from different experimental settings and are not directly interchangeable. 

\textbf{$K_d$ (dissociation constant)} is a thermodynamic parameter that measures how tightly a ligand binds its target, typically expressed in molarity (e.g., nM). In contrast, \textbf{$IC_{50}$} (half maximal inhibitory concentration) represents the functional potency of a molecule in neutralization or inhibition assays, often reported in mass concentration units (e.g., µg/mL), and is influenced by assay-specific factors such as target expression levels, cell type, or incubation time.

Crucially, a high-affinity binder (low $K_d$) does not always translate to strong inhibition (low $IC_{50}$), and vice versa. For instance, an antibody may bind its antigen tightly but at an epitope that is not functionally relevant (e.g., non-neutralizing sites), resulting in poor inhibitory effect. Conversely, moderate-affinity antibodies may achieve strong neutralization if they block a key functional domain, such as a receptor-binding site or active cleft. This biological dissociation between affinity and function underscores the importance of evaluating both types of measurements when studying binding prediction.

Under simplified assumptions, the two are often treated as approximately comparable, particularly when no competing ligands are involved. The theoretical relationship can be approximated using the Cheng–Prusoff equation:

\[
K_d = \frac{IC_{50}}{1 + \frac{[S]}{K_m}},
\]

which simplifies to $K_d \approx IC_{50}$ in the absence of competing substrates or saturating conditions.

\paragraph{Unit Conversion:}
To compare $IC_{50}$ values (typically in µg/mL) with $K_d$ (typically in nM), we convert $IC_{50}$ to molar units using the molecular weight ($MW$) of the antibody:

\[
IC_{50}^{\text{(nM)}} = \frac{IC_{50}^{\text{(µg/mL)}} \times 10^3}{MW^{\text{(Da)}}}.
\]

A typical IgG antibody has a molecular weight of approximately \textbf{150 kDa} (i.e., 150,000 Da), yielding:

\[
1\ \mu\text{g/mL} \approx \frac{1000}{150,000} = 6.67\ \text{nM}.
\]

Thus, when converting $IC_{50}$ to molar units, we assume a reference molecular weight of 150 kDa unless otherwise stated.

\newpage

\section{Baseline Implementation Details}
\label{SI:baseline}


\textbf{IgFold}~\citep{ruffolo2023fast}: We utilized IgFold for structure prediction of antibody structures, employing the official open-source implementation available at \url{https://github.com/Graylab/IgFold}. We skipped the refinement step to output backbone-only structures. 

\textbf{Boltz-1}~\citep{wohlwend2024boltz}: We generated both single Ags and Ab-Ag complex structures using Boltz-1, an open-source biomolecular structure prediction tool available at \url{https://github.com/jwohlwend/boltz}. Our implementation utilized the MSA-based inference pipeline, where multiple sequence alignments (MSAs) were generated using ColabFold's MMseqs2-based search tool with default parameters. These MSAs were then input into Boltz-1 to predict 3D structures of antibody–antigen complexes and single antigens.

\textbf{AlphaFold 3 (AF3)}~\citep{abramson2024accurate}: Structures were predicted using AlphaFold 3 via a local installation equipped with NVIDIA H100 GPUs. We used the official implementation available at \url{https://github.com/google-deepmind/alphafold3}. Structure generation followed the default model configuration provided in the repository. For each complex, we selected the top-ranked prediction based on the pLDDT confidence score.

\textbf{MODELLER}~\citep{vsali1993comparative}: We used MODELLER 10.6 (\url{https://salilab.org/modeller/}) to generate complex structures for single amino acid mutants, using wild-type Ab–Ag complexes predicted by AF3 or Boltz-1 as templates.

-----------------------------------------------------------------------------------------------------------------------

\textbf{Mint}~\citep{ullanat2025learning}: We utilized the official MINT implementation (\url{https://github.com/VarunUllanat/mint}) to embed antibody–antigen complexes. For each complex, the full sequence—including both heavy and light antibody chains along with the antigen—was concatenated and input directly into the model, as clarified in the official repository (\url{https://github.com/VarunUllanat/mint/issues/7}). This approach aligns with MINT's design, which processes the entire complex sequence holistically. To predict binding affinity, we appended a multi-layer perceptron (MLP) consisting of two hidden layers with ReLU activations and a dropout rate of 0.2. The model was trained jointly using a mean squared error (MSE) loss for regression and a pairwise margin ranking loss with a margin of 0.1. Optimization was performed using the Adam optimizer with a learning rate of 1e-4, and early stopping based on validation loss was employed over 100 training epochs.

\textbf{ANTIPASTI}~\citep{michalewicz2024antipasti}: We used the ANTIPASTI code from the official GitHub repository (\url{https://github.com/kevinmicha/ANTIPASTI}). The model was run in antibody–antigen interface prediction mode, and embeddings from the penultimate layer were extracted as features. These were passed to a downstream regression head trained on our binding affinity targets.

\textbf{GearBind}~\citep{cai2024pretrainable}: The Gearbind code was obtained from its GitHub repository (\url{https://github.com/DeepGraphLearning/GearBind}). We followed the default protocol for feature extraction and used the pre-trained sequence–structure binding classifier. To accommodate insertions and deletions in our sequences, we applied custom preprocessing steps to align the input format with GearBind’s requirements. The resulting binding scores were used directly as proxies for affinity.

\textbf{PBEE}~\citep{chaves2025estimating}: We used the PBEE model from its official GitHub (\url{https://github.com/chavesejf/PBEE}). Structures were parsed into the required input format using the preprocessing scripts provided by the authors. PBEE was run with its default configuration to score complexes, and predicted values were interpreted as binding strength estimates.

\textbf{FoldX 5.1}~\citep{delgado2019foldx}: FoldX calculations were performed using FoldX version 5.1 under an academic license (\url{https://foldxsuite.crg.eu/}). For each antibody–antigen complex, structures were first repaired using the \texttt{RepairPDB} command to correct potential structural issues. We then ran \texttt{AnalyseComplex} to estimate the interaction energies between subunits. Specifically, we summed the interaction energies of the antigen–heavy chain (Ag–AbH), antigen–light chain (Ag–AbL), and heavy–light chain (AbH–AbL) interfaces to obtain the total binding energy (\(\Delta G\)), which was used as a proxy for binding affinity. All FoldX runs used default parameters.

-----------------------------------------------------------------------------------------------------------------------

\newpage

\section{Supplementary figures}

\begin{figure}[h!t]
\centering
    \captionsetup{width=1\linewidth}
    \includegraphics[width=1\linewidth]{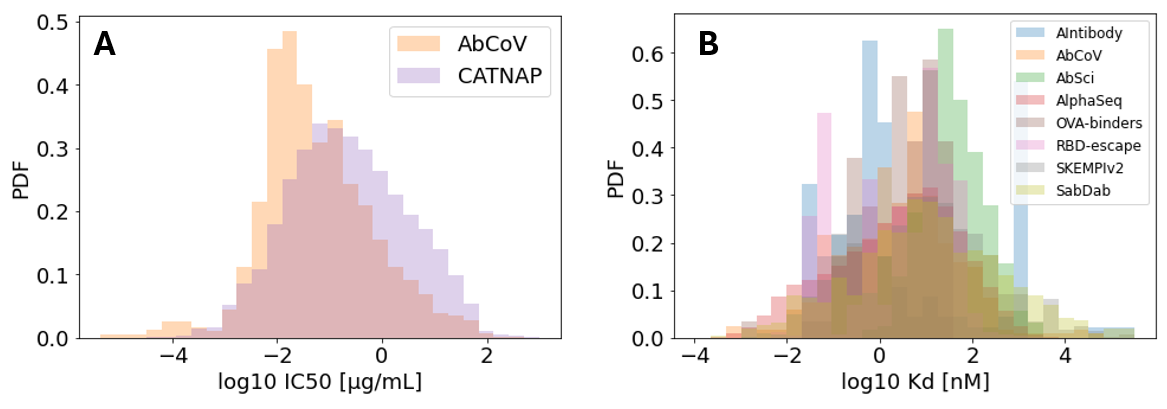}
    \caption{Probability density function (PDF) of binding affinity values across source databases. Some sources report (A) $IC_{50}$ values, while others report (B) $K_d$ values.}
    \label{fig:database_affinity}
\end{figure}

\newpage

\section{Supplementary Tables}


\def\arraystretch{1.3}
\begin{table}[ht]
    \centering
    \caption{License information for each individual datasets included in AbRank.\\ *The Kaggle dataset will be updated when AIntibody becomes publicly available. }
    \vspace{0.2cm}
    \begin{tabular}{p{3cm} l}
    \hline
    \textbf{Dataset} & \textbf{License} \\
    \hline
    HER2-binders & Clear BSD \\
    \hline
    SAbDab & CC BY 4.0 \\
    \hline
    SKEMPI 2.0 & CC BY 4.0 \\
    \hline
    AlphaSeq & CC BY 4.0 (via Zenodo) \\
    \hline
    CATNAP & CC BY 4.0 (Open Access via \textit{Nucleic Acids Research}) \\
    \hline
    Ab-CoV & CC BY 4.0 (Open Access via \textit{Bioinformatics}) \\
    \hline
    OVA-binders & CC BY 4.0 (Open Access via \textit{Communications Biology}) \\
    \hline
    RBD-escape & GPL v3 (\url{https://github.com/jbloomlab/SARS2_RBD_Ab_escape_maps}) \\
    \hline
    AIntibody & Not publicly released yet* (\url{https://www.aintibody.org/})\\
    \hline
    \end{tabular}
    \label{tab:dataset_licenses}
\end{table}
\def\arraystretch{1.0}

\def\arraystretch{1.2}
\begin{table}[h]
    \centering
    \captionsetup{width=1\textwidth}
    \caption{Dataset statistics for each benchmark split, including the number of unique Ab–Ag complexes with affinity values and ranking pairs involved.}
    \vspace{0.1cm}
    \label{tab:dataset-count}
    \begin{tabular}{lll}
    \toprule
    \textbf{Benchmark Split} & \textbf{Number of complexes with affinity values} & \textbf{Number of Pairs considered} \\
    \midrule
    Training - Balanced & 61,612 & 168,470 \\
    Training - Hard Ab & 42,959 & 130,852 \\
    Training - Hard Ag & 1,165 & 3,470 \\
    Test - Unrelated Complex & 362 & 1200 \\
    Test - Local Perturbation & 362 & 706 \\
    \bottomrule
    \end{tabular}
\end{table}
\def\arraystretch{1.0}

\end{document}